\documentclass[draft]{agujournal2019}
\usepackage{url} %this package should fix any errors with URLs in refs.
\usepackage{lineno}
\usepackage{amsfonts}
\usepackage[inline]{trackchanges} %for better track changes. finalnew option will compile document with changes incorporated.
\usepackage{soul}
\draftfalse

\journalname{Geophysical Research Letters}

\begin{document}

\title{Radial Diffusion Driven by Spatially Localized ULF Waves in the Earth's Magnetosphere.}
 
\authors{Adnane Osmane\affil{1}, Jasmine K. Sandhu\affil{2}, Tom Elsden\affil{3}, Oliver Allanson\affil{4, 5,6} and Lucile Turc\affil{1}}

\affiliation{1}{University of Helsinki, Department of Physics, Helsinki, Finland}
 \affiliation{2}{University of Leicester, Department of Physics and Astronomy, Leicester, UK}
 %\affiliation{3}{Northumbria University, Department of Mathematics, Physics and Electrical Engineering, Newcastle Upon Tyne, UK}
\affiliation{3}{University of St Andrews, School of Mathematics and Statistics, St. Andrews, UK}
\affiliation{4}{University of Birmingham, Space Environment and Radio Engineering, School of Engineering, Birmingham, UK}
\affiliation{5}{University of Exeter, Environmental Mathematics, Department of Earth and Environmental Sciences, Penryn,UK}
\affiliation{6}{University of Exeter, Department of Mathematics, Exeter, UK}

\correspondingauthor{Adnane Osmane}{adnane.osmane@helsinki.fi}

\begin{keypoints}
\item A radial diffusion coefficient for spatially localized ULF waves is derived for the first time.
\item ULF waves sampled on less than 10\% of the drift orbit provide more efficient diffusion by a factor of 10 to 25 \%.
\item Our results apply to particles transiting through magnetospheric plumes and MLT localised ULF wave packets.
\end{keypoints}

%\twocolumn
\begin{abstract}
Ultra-Low Frequency (ULF) waves are critical drivers of particle acceleration and loss in the Earth's magnetosphere. While statistical models of ULF-induced radial transport have traditionally assumed that the waves are uniformly distributed across magnetic local time (MLT), decades of observational evidence show significant MLT localization of ULF waves in the Earth's magnetosphere. This study presents, for the first time, a quasi-linear radial diffusion coefficient accounting for localized ULF waves. We demonstrate that even though quasi-linear radial diffusion is averaged over drift orbits, MLT localization significantly alters the efficiency of particle transport. Our results reveal that when ULF waves cover more than 30\% of the MLT, the radial diffusion efficiency is comparable to that of uniform wave distributions. However, when ULF waves are confined within 10\% of the drift orbit, the diffusion coefficient is enhanced by 10 to 25\%, indicating that narrowly localized ULF waves are efficient drivers of radial transport.
\end{abstract}

%\section*{Plain Language Summary}
%Enter your Plain Language Summary here or delete this section.
%Here are instructions on writing a Plain Language Summary: 
%https://www.agu.org/Share-and-Advocate/Share/Community/Plain-language-summary

%%%%%%%%%%%%%%%%%%%%%%%%%%%%%%%%%%%%%%%%%%%%%%%
%
%  BODY TEXT
%
%%%%%%%%%%%%%%%%%%%%%%%%%%%%%%%%%%%%%%%%%%%%%%%

\section{Introduction} 
Ultra-Low Frequency waves are key drivers of particle transport, acceleration, and losses in the Earth’s radiation belts \cite{su2015ultra,Jaynes18, George22b, olifer2024rapid}. For decades, statistical models of ULF-induced radial transport, typically formulated using Fokker-Planck equations, have assumed that the waves are uniformly distributed across magnetic local time (MLT) \cite{Falthammar65, Elkington99, Elkington03, Lejosne19, Osmane21a, osmane2023radial}. However, decades of observational evidence have shown that ULF waves are often significantly localized in MLT \cite{Murphy20} due to localised source regions and inhomogeneous properties of the magnetospheric plasma. For example, storm-time Pc5 waves with high azimuthal wave numbers, typically in the poloidal mode, are driven by unstable ring current ion distributions and substorm injections, and are predominantly observed in the pre-midnight sector \cite{Anderson90}. Statistical studies have further demonstrated that Pc5 fluctuations exhibit strong MLT-dependent variations in wave power and radial distance, as well as pronounced day-night asymmetries—wave power is stronger on the nightside in the inner magnetosphere and on the dayside near the magnetopause \cite{Liu09,Sarris22,Yan23}. Additionally, toroidal field line resonances show MLT dependence in amplitude, with larger amplitudes observed on the dawn flank compared to the dusk flank. This pattern has been widely observed both on the ground \cite{Gupta75,Vennestrom99,Pahud09,Rae12} and in situ \cite{Anderson90,Kokubun13,Takahashi15,Takahashi16,Yan23}. 

Despite extensive observational data highlighting MLT-dependent wave localisation, the impact on radial diffusion coefficients $D_{LL}$ remains unknown. Since radial diffusion typically tracks the drift-averaged evolution of the particle distribution function, it is unclear whether the MLT-localized ULF waves are more or less efficient in driving radial transport compared to waves uniformly distributed in MLT. This paper bridges the gap between modeling and observations by deriving, for the first time, a quasi-linear radial diffusion coefficient for MLT-localized ULF waves. In this study, we address two key questions: (1) What is the parametric dependence of the radial diffusion coefficient on MLT localization of ULF waves? and (2) Is radial diffusion more or less efficient when ULF waves are spatially localized compared to when they are uniformly distributed across MLT?

\begin{figure}
  \label{Cartoon}
  %\centering
  \includegraphics[width=0.50\textwidth]{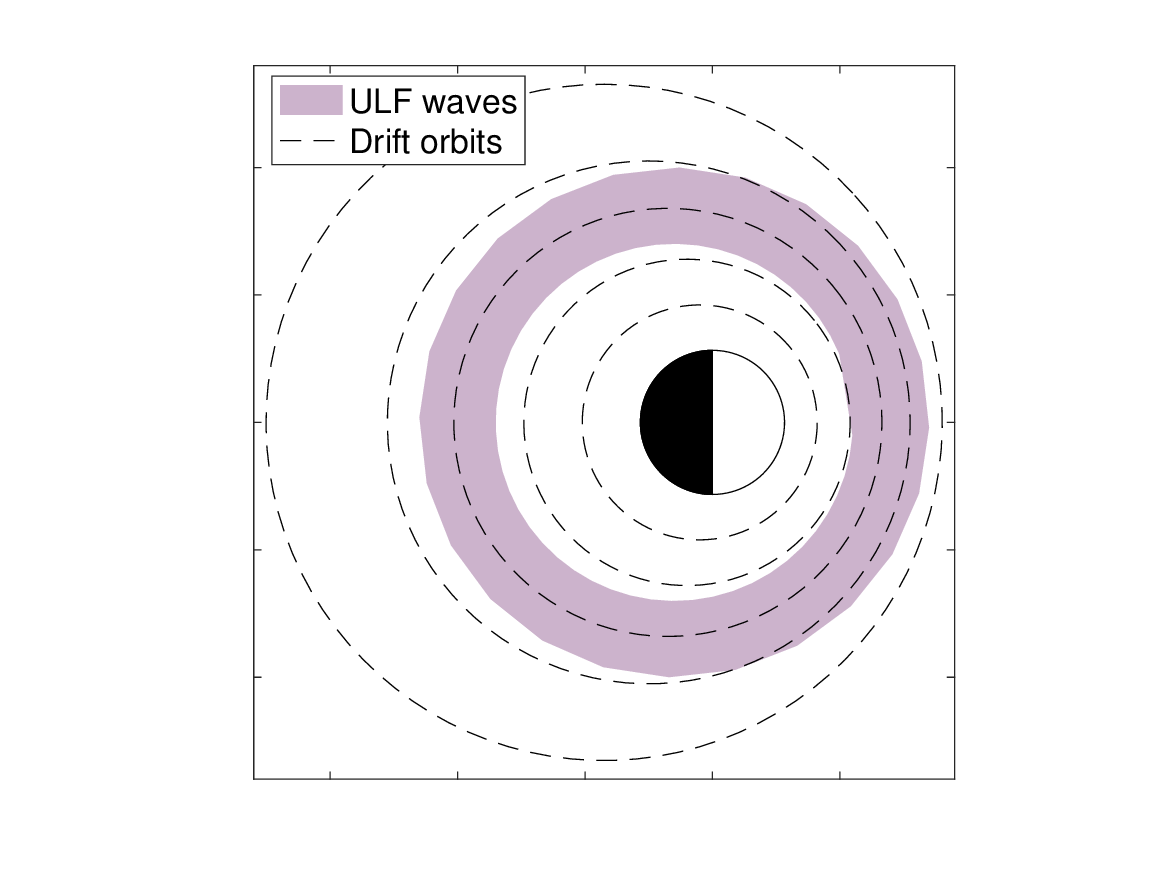}
    \includegraphics[width=0.50\textwidth]{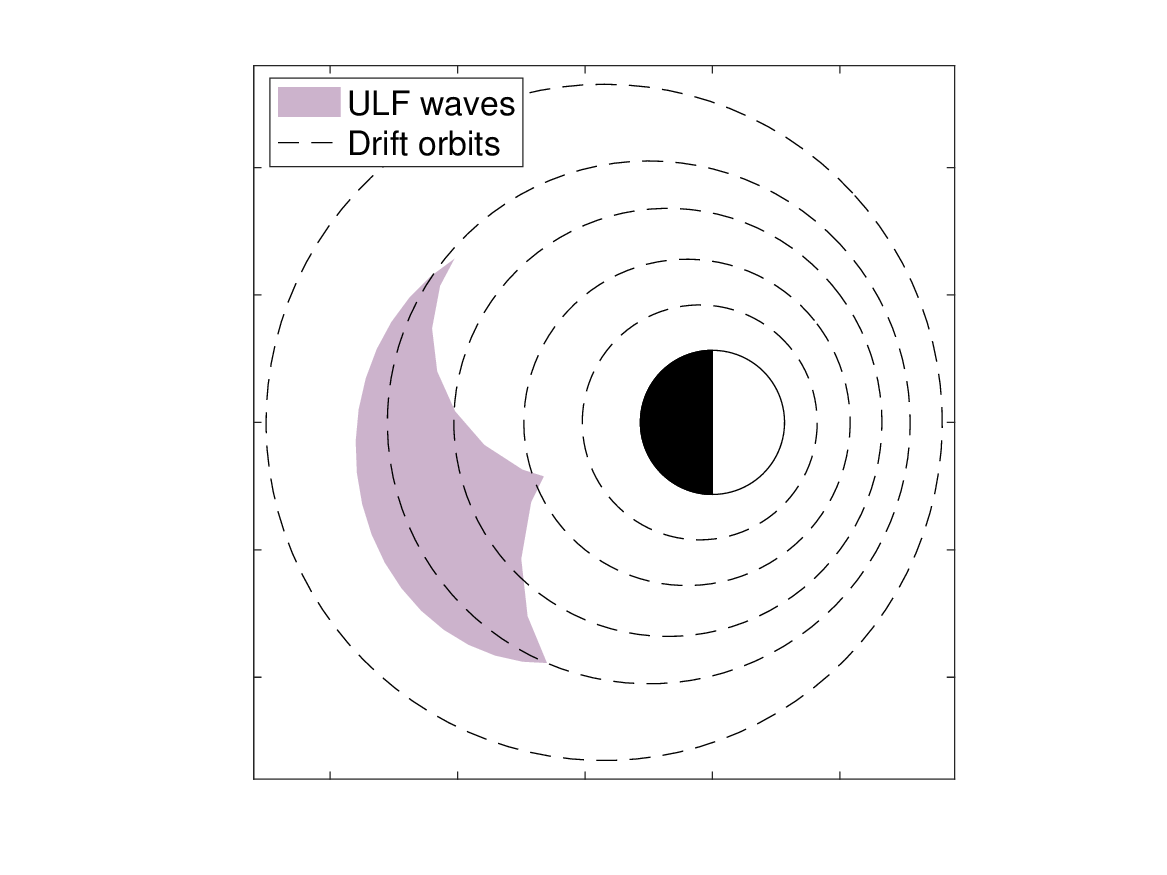}
 \caption{Illustration of two general scenarios where magnetically trapped particles encounter MLT-localized ULF waves. In the left panel, the ULF wave is homogeneous across MLT but localized in radial distance. As a result, particles on different drift shells move in and out of the plasma regions where the ULF waves are present. In the right panel, the ULF waves are localized in MLT, meaning that all trapped particles only encounter the waves within specific MLT regions.}
 \end{figure}

In the following, we consider two scenarios under which magnetically trapped particles in the Earth's magnetosphere can encounter MLT localised ULF waves. These two scenarios are illustrated in Figure (\ref{Cartoon}). In the first scenario, shown in the left panel, the ULF wave is homogeneous across MLT but localized in radial distance. As a result, particles on different drift shells move in and out of the plasma regions where the ULF waves are present. In the second scenario, shown in the right panel, the ULF waves are localized in MLT, meaning that all trapped particles only encounter the waves within specific MLT regions. This second case may occur due to magnetospheric plumes or regions where ULF waves are locally generated in the magnetosphere due to boundary instabilities, e.g., Kelvin-Helmholtz instabilities at the magnetopause \cite{masson2018kelvin}, the impact of upstream mesoscale transient structures \cite{Hartinger13, Kajdic24}, or particle driven drift or bounce resonant instabilities \cite{Southwood81}. 

Although from the particle’s perspective, the distinction between MLT localization due to Scenario 1 (left panel of Figure \ref{Cartoon}) or Scenario 2 (right panel of Figure \ref{Cartoon}) is irrelevant—since the particle only samples the field along its trajectory—it is important to emphasize that MLT-localized ULF waves might be the norm rather than the exception when treating wave-particle interactions. ULF waves are often radially and MLT-localized, and this spatial localization occurs independently of the drift orbits confined to specific drift shells.  The calculations presented here can thus be applied to a wide range of geomagnetic conditions, including both storm and non-storm periods, where ULF waves exhibit spatial localization. 

This communication is organized as follows. In Section \ref{Methodology}, we first describe the electromagnetic field model that accounts for MLT-localized ULF waves (Section \ref{ULF_wave_model}), followed by a step-by-step quasi-linear derivation of the associated radial diffusion coefficient $D_{LL}$ (Section \ref{QLT_derivation}). We then compare our results to the general case where ULF waves are uniformly distributed in MLT and provide a physical explanation for the observed discrepancies when ULF waves are MLT-localized (Section \ref{explanation}). Finally, in Section \ref{conclusion}, we summarize our findings, discuss the key limitations of our work, and outline future steps to further quantify the impact of ULF waves on particle transport.

\section{Methodology}\label{Methodology}
\subsection{ULF Wave Model}\label{ULF_wave_model}
We proceed with a derivation of radial diffusion for the special case of an electrostatic poloidal electric field superposed onto a dipolar magnetic field. The background magnetic field is given by: 
%%%%%%%%%%%%%%%%%%%%%%%%%%%%%%%%
\begin{equation}
\label{Magnetic_field_eq_1}
    \mathbf{B}=-\frac{B_E R_E^3}{r^3} \hat{\mathbf{z}},
\end{equation}
%%%%%%%%%%%%%%%%%%%%%%%%%%%%%%%%
in terms of the magnetic moment $B_E$ and the Earth's radius $R_E\simeq 6300$ km. We can ignore the latitude dependence of the background magnetic field because we derive radial diffusion for the special case of equatorially trapped particles, and thus the unit magnetic field vector $\mathbf{b}=\frac{\mathbf{B}}{|B|}=-\hat{\mathbf{z}}$ always points in the negative z direction in a cyclindrical system of coordinates written in terms of $(r, \varphi, z)$, where $\varphi$ is the azimuthal angle. The poloidal electric field due to the ULF wave is written as: 
%%%%%%%%%%%%%%%%%%%%%%%%%%%%%%%%
\begin{eqnarray}
\label{electric_field_eq2}
\mathbf{E}&=&   {\delta E}_\varphi (r, \varphi, t) \hat{\varphi} \nonumber \\
&=& f(\varphi; \kappa) \sum_m {\delta E}_{\varphi, m}\frac{R_E}{r} e^{im\varphi}  {\hat{\varphi}}, 
\end{eqnarray}
%%%%%%%%%%%%%%%%%%%%%%%%%%%%%%%%
in terms of the time-dependent Fourier coefficients $\delta E_{\varphi, m} (t)$ and the von Mises distribution:
%%%%%%%%%%%%%%%%%%%%%%%%%%%%%%%%
\begin{eqnarray}\label{Eq:vonMises}
    f(\varphi; \kappa)&=&\frac{e^{\kappa \cos(\varphi)}}{2\pi I_0(\kappa)}, \nonumber\\
&=&\frac{1}{2\pi}\left(1+2\sum_{n=1}^\infty \psi_n(\kappa) \cos(n\varphi)\right)   
\end{eqnarray}
%%%%%%%%%%%%%%%%%%%%%%%%%%%%%%%%
the latter of which is written in terms of the modified Bessel function of the first kind with degree zero $I_0(\kappa)$ \cite{abramowitz1968handbook}, and the parameter $\kappa \in [0, \infty )$. The von Mises distribution is the analogue to the Maxwellian distribution on a circle and normalised. Written in the above form, it is also normalised as can be shown when integrated along $\varphi$ between $-\pi$ and $\pi$. We note that this representation has previously been used to describe coherent interactions between ULF waves and magnetically confined electrons in the Earth's radiation belts \cite{Li18, hao2020simultaneous} and that other representations should provide the same transport coefficients, with the stipulation discussed in Section (\ref{QLT_derivation}) and derived in \ref{Appendix_Fourier} to constrain the root mean square fluctuations sampled over a drift orbit.

 % \begin{figure}
  %\label{Fig_2}
 % \centering
% \includegraphics[width=0.6\textwidth]{Fig1a.eps}
% \caption{The von Mises distribution function is parameterized for $\kappa$ values of $[0, 1.77, 3.3, 8.6, 30]$. When $\kappa = 0$, the distribution in magnetic local time (MLT) is uniform. As $\kappa$ increases, the distribution becomes progressively narrower around the mean value of $\varphi = 0$. The selected $\kappa$ values correspond to cases where the envelope confines ULF waves to approximately 100, 75, 50, 30 and 15 percent of the drift orbit.}
 %\end{figure}

 The choice of the electric field dependence on the radial distance and the azimuthal angle requires an explanation. The term inside the sum of Equation (\ref{electric_field_eq2}) is a Fourier decomposition for coefficients $\delta {E}_{\varphi, m} R_E/r$. The $1/r$ dependence of the electric field comes from Faraday's equation for an electrostatic field that satisfies $\nabla \times \mathbf{E}=0$. In cylindrical coordinates, we are required to keep $\delta E_\varphi r$ constant. The introduction of the von Mises distribution in Equation (\ref{electric_field_eq2}) allows us to confine the ULF electric field to any range of magnetic local times we deem appropriate. ULF modes of azimuthal wave numbers $m$ are also multiplied by the envelope $f(\varphi; \kappa)$ to confine the mode within a localised MLT region. The range of MLT localisation is parameterised by the coefficient $\kappa$. If $\kappa=0$, the ULF field is uniformly spread across all magnetic local times. But as $\kappa$ becomes much greater than one, the von Mises distribution converges to a Maxwellian: 
%%%%%%%%%%%%%%%%%%%%%%%%%%%%%%%%
\begin{equation}
   \lim_{\kappa \longrightarrow \infty} f(\varphi;\kappa) = \sqrt{\frac{\kappa}{2\pi}} e^{-\kappa \varphi ^2/2}. 
\end{equation}   
%%%%%%%%%%%%%%%%%%%%%%%%%%%%%%%%
Thus, the parameter $\kappa$ is equivalent to the inverse of the variance of a Maxwellian, and the von Mises distribution appearing in the Fourier decomposition of the poloidal electric field in Equation (\ref{electric_field_eq2}) confines the bulk of the ULF fluctuations around $\varphi= \pm 2 \sqrt{1/\kappa}$. 

In the following, we use a Fourier representation of the poloidal electric field derived from Equation (\ref{electric_field_eq2}). The derivation is provided in \ref{Appendix_Fourier} and given by the following decomposition: 
%%%%%%%
\begin{eqnarray}
{E_\varphi}&=& \frac{R_E}{r}\sum_m c_m e^{im\varphi}, 
\end{eqnarray}
%%%%%%%
with the Fourier coefficients: 
%%%%%%%
\begin{eqnarray}
    c_m=\sum_{m'} \frac{{{\delta E}}_{\varphi, m'}}{2 \pi} \left(\delta_{m, m'}+2\varepsilon_{m,m'}\psi_{m'-m}\right),
\end{eqnarray}
%%%%%%%
where we used the Kronecker delta  $\delta_{m, m'}$ for indices $m$ and $m'$ and its complement $\varepsilon_{m,m'}=1-\delta_{m,m'}$. The coefficient $\psi_n$ is a ratio of the modified Bessel function of degree $n$ to the modified Bessel function of degree $0$: 
\begin{equation}
\psi_n(\kappa) = \frac{I_n(\kappa)}{I_0(\kappa)}.
\end{equation}
Note that the modified Bessel function of the first kind is symmetric with respect to a transformation of the coefficients $n \in \mathbb{Z}$ with $I_{-n}(\kappa) = I_n(\kappa)$, and thus $\psi_{-n}=\psi_n$. 
With the electric and magnetic field given by Equations (\ref{Magnetic_field_eq_1}) and (\ref{electric_field_eq2}), we can compute the guiding center drift velocities \cite{Hazeltine18}. For particles trapped in the equatorial plane, we only have two drift contributions, the $\mu \nabla B$ drift, for a particle of charge $q$, relativistic Lorentz factor $\gamma$ and first adiabatic invariant $\mu$: 
%%%%%%%%%%%%%%%%%%%%%%%%%%%%%%%%
\begin{equation}\label{mugradBdrift}
    -\frac{\mu\nabla B\times \mathbf{b}}{q\gamma B} =\frac{3 \mu}{q\gamma r} \hat{\varphi}
\end{equation}
%%%%%%%%%%%%%%%%%%%%%%%%%%%%%%%%
and the $E$ cross $B$ drift: 
%%%%%%%%%%%%%%%%%%%%%%%%%%%%%%%%
\begin{eqnarray}\label{EcrossBequation8}
\frac{\mathbf{E} \times \mathbf{b}}{B}&=&-\hat{r}\sum_m \frac{c_m}{B_E} \frac{r^2}{R_E}e^{im\varphi} \\ 
&=&-\hat{r} \sum_{m}\sum_{m'}\frac{1}{2\pi}\frac{\delta E_{\varphi, m'}}{ B_E}\frac{r^2}{R_E}  \left(\delta_{m, m'}+2\varepsilon_{m,m'}\psi_{m'-m}\right)e^{im\varphi}. 
\end{eqnarray}
%%%%%%%%%%%%%%%%%%%%%%%%%%%%%%%%
The drift velocity $\mathbf{v}_d$ can therefore be written as: 
%%%%%%%%%%%%%%%%%%%%%%%%%%%%%%%%
\begin{equation}\label{drift_velocity}
    \mathbf{v}_d=\frac{3 \mu}{q\gamma r} \hat{\varphi}
-\hat{r} \sum_{m}\sum_{m'}\frac{1}{2\pi}\frac{\delta E_{\varphi, m'}}{ B_E}\frac{r^2}{R_E}  \left(\delta_{m, m'}+2\varepsilon_{m,m'}\psi_{m'-m}\right)e^{im\varphi}
\end{equation}
%%%%%%%%%%%%%%%%%%%%%%%%%%%%%%%%
The radial diffusion equation can be derived from the drift velocity (\ref{drift_velocity}) using several approaches. One common method involves calculating the increment $\Delta r$ from Equation (\ref{drift_velocity}), applying standard quasi-linear assumptions that the electric field amplitude $\delta E_{\varphi, m}$ behaves as white noise, and then computing the drift average of $\Delta r^2$ to estimate the diffusion coefficient (see, e.g., \citeA{Falthammar65, Elkington03, Fei06, Lejosne19}). However, in this work, we derive the quasi-linear radial diffusion coefficient directly from the kinetic equation. This approach, previously applied by \citeA{osmane2023radial} for ULF waves and by \citeA{Kennel66} for kinetic scale fluctuations, is mathematically more elaborate but offers more clarity in how quasi-linear assumptions are incorporated into the derivation and how higher-order and non-diffusive effects are omitted.

\subsection{Kinetic Equation and Quasi-Linear Derivation}\label{QLT_derivation}
In the absence of loss processes, the kinetic equation for equatorially trapped guiding centers in planetary radiation belts (see \citeA{osmane2023radial} and references therein) can be expressed in conservative form as:
%%%%%%%%%%%%%%%%%%%%%%%%%%%%%%%%
\begin{equation}\label{kinetic1}
  \frac{\partial}{\partial t}  (B g)+\nabla\cdot (\mathbf{v}_d Bg)=0.
\end{equation}
%%%%%%%%%%%%%%%%%%%%%%%%%%%%%%%%
The magnetic field amplitude B is included in Equation (\ref{kinetic1}) because it represents the Jacobian of the transformation when the particle’s velocity is expressed in field-aligned coordinates. The distribution function of guiding centers is denoted as  $g=g(r, \varphi, t)$. Utilizing Liouville’s theorem
%%%%%%%%%%%%%%%%%%%%%%%%%%%%%%%%
\begin{equation}
     \frac{\partial}{\partial t}  (B)+\nabla\cdot (\mathbf{v}_d B)=0,
\end{equation}
%%%%%%%%%%%%%%%%%%%%%%%%%%%%%%%%
the evolution of the distribution function is given by: 
%%%%%%%%%%%%%%%%%%%%%%%%%%%%%%%%
\begin{equation}\label{kinetic_2}
\frac{\partial g}{\partial t} +\frac{3 \mu}{q\gamma r^2} \frac{\partial g}{\partial \varphi}=\sum_m \frac{c_m}{B_E}\frac{r^2}{R_E^2}e^{im\varphi} \frac{\partial g}{\partial r}. 
\end{equation}
%%%%%%%%%%%%%%%%%%%%%%%%%%%%%%%%
Note the appearance of the azimuthal drift frequency $\Omega_d=3\mu /q\gamma r^2$ on the left-hand side of Equation (\ref{kinetic_2}). 

The next step is to decompose the distribution function into two components: one part, denoted as $\delta g$, that evolves on fast timescales comparable to or shorter than the drift period, and another part that evolves on much longer timescales, surpassing both the drift period and the characteristic timescales of Pc4 and Pc5 ULF wave frequencies:
%%%%%%%%%%%%%%%%%%%%%%%%%%%%%%%%
\begin{eqnarray}\label{decomposition_g}
    g(r, \varphi, t)&=&g_0(r, t) + \delta g (r,\varphi, t) \nonumber  \\ 
        &=& g_0(r, t) + \sum_p \delta g_p(r, t) e^{ip\varphi}. 
\end{eqnarray}
%%%%%%%%%%%%%%%%%%%%%%%%%%%%%%%%
In Equation (\ref{decomposition_g}), the fast part of the distribution function, which is function of the azimuthal angle $\varphi$, is decomposed in terms of Fourier modes to simplify our analysis. The slow part of the distribution can be extracted by the following averaging procedure: 
\begin{equation}\label{averaging_procedure}
    g_0(r, t)=\langle g \rangle_\varphi=\frac{1}{2\pi T} \int_{-\pi}^{\pi} \mathrm{d}\varphi \int_0^T \mathrm{d}t \hspace{1mm}g(r, \varphi, t). 
\end{equation}
Inserting the decomposition of Equation (\ref{decomposition_g}) inside the kinetic equation, multiplying by a factor $e^{-iq\varphi}$, integrating from $-\pi$ to $\pi$, and keeping the terms for the index $q=0$ (i.e. keeping only the MLT averaged part), we find the following evolution equation for $g_0$: 
%%%%%%%%%%%%%%%%%%%%%%%%%%%%%%%%
\begin{equation}\label{Equation_slow_part}
    \frac{\partial g_0}{\partial t}=\frac{r^2}{R_E^2} \frac{c_{q=0}}{B_E} \frac{\partial g_0}{\partial r}+\frac{r^2}{R_E^2}\sum_m\frac{\partial }{\partial r}\left(\frac{c^*_m \delta g_m}{B_E}\right)
\end{equation}
%%%%%%%%%%%%%%%%%%%%%%%%%%%%%%%%
The terms $c_{m}^*$ in Equation (\ref{Equation_slow_part}) represent the complex conjugate of the electric field Fourier coefficient and arises after inversing the sum over $m$ with $m\longrightarrow -m$ because the electric field is a real quantity, leading to the conditions $c_{-m}=c_{m}^*$. 

The quasi-linear equation that tracks transport on timescales longer than the drift period can be retrieved by applying the averaging procedure (\ref{averaging_procedure}) on Equation (\ref{Equation_slow_part}): 
%%%%%%%%%%%%%%%%%%%%%%%%%%%%%%%%
\begin{equation}\label{Equation_slow_part2}
\frac{\partial g_0}{\partial t}=\sum_m \frac{1}{B_E}\frac{r^2}{R_E^2}\frac{\partial}{\partial r} \bigg{(}\langle c_{m}^*\delta g_{m} \rangle_\varphi \bigg{)}
\end{equation}
%%%%%%%%%%%%%%%%%%%%%%%%%%%%%%%%
We note that the first term on the right-hand side of Equation (\ref{Equation_slow_part}) cancels after averaging on timescales comparable to the ULF wave period since $\langle \delta E_{\varphi, m=0}\rangle_\varphi =0$. 

In order to solve the quasi-linear Equation (\ref{Equation_slow_part2}), we need an equation for the perturbed part of the distribution function $\delta g_m$. The procedure is identical to the one outlined in Appendix B of \citeA{osmane2023radial} and results in the following equation: 
%%%%%%%%%%%%%%%%%%%%%%%%%%%%%%%%
\begin{eqnarray}\label{Perturbed_part}
    \frac{\partial \delta g_m}{\partial t} + \underbrace{im \Omega_d \delta g_m}_{\textrm{\small{Azimuthal drifting}}}&=&\underbrace{\frac{r^2}{R_E^2}\frac{c_m}{B_E} \frac{\partial g_0}{\partial r}}_{\textrm{\small{Linear wave-particle interaction}}}
    +\underbrace{\frac{r^2}{R_E^2}\sum_{m'} \frac{c_{m'}^*}{B_E} \frac{\partial \delta g_{m+m'}}{\partial r}}_{\textrm{\small{Higher order wave-particle interaction}}}
\end{eqnarray}
%%%%%%%%%%%%%%%%%%%%%%%%%%%%%%%%
In the quasi-linear regime, the perturbed part of the distribution function is assumed to evolve linearly. Consequently, higher-order wave-particle interactions in Equation (\ref{Perturbed_part}) are neglected, reducing the problem to a linear equation with the following general solution:
%%%%%%%%%%%%%%%%%%%%%%%%%%%%%%%%
\begin{equation}\label{linear_solution}
  \delta g_m(r, t)=\delta g_m(r,0)e^{-im\Omega_d t}+\frac{1}{B_E}\frac{r^2}{R_E^2}\int_0^t  \mathrm{d}t'\   c_{m}(t') e^{-im\Omega_d (t'-t)}\frac{\partial g_0}{\partial r} 
\end{equation}
%\begin{equation}\label{linear_solution}
%  \delta g_m(r, t)=\delta g_m(r,0)e^{-im\Omega_d t}+\sum_{m'}2\psi_{m'-m}\frac{r^2}{R_E^2}\int_0^t  \mathrm{d}t'\   A_{m', \varphi}(t') e^{-im\Omega_d (t'-t)}\frac{\partial g_0}{\partial r} 
%\end{equation}
%%%%%%%%%%%%%%%%%%%%%%%%%%%%%%%%
The first term on the right-hand side of Equation (\ref{linear_solution}) generates drift echoes, and since over long diffusive time, drift echoes experience phase-mixing, we also ignore their contribution to the evolution of $g_0$ and set $\delta g_m(r, 0)=0$. 

Inserting Equation (\ref{linear_solution}) into Equation (\ref{Equation_slow_part2}) we find: 
%%%%%%%%%%%%%%%%%%%%%%%%%%%%%%%%
\begin{eqnarray}\label{Equation_diffusion_1}
\frac{\partial g_0}{\partial t}&=&\sum_m \frac{1}{B_E^2}\frac{r^2}{R_E^2}\frac{\partial}{\partial r} \bigg{[}
\frac{r^2}{R_E^2}\int_0^t  \mathrm{d}t'\  \bigg{\langle} c_{m}(t')c_{m}^*(t) \bigg{\rangle}_\varphi e^{-im\Omega_d (t'-t)}\frac{\partial g_0}{\partial r} \bigg{]} \nonumber 
\end{eqnarray}
%%%%%%%%%%%%%%%%%%%%%%%%%%%%%%%%
The above diffusive equation is a function of the correlator $\big{\langle} c_{m}(t)c_{m}^*(t') \big{\rangle}_\varphi$ for the ULF wave electric field. Within quasi-linear theories, we assume that the fluctuations driving the system are akin to stationary white noise, i.e., $\langle c_m(t) c_{m}^*{t'}\rangle \simeq \delta(t'-t)$, and that the background distribution function $g_0$ is not significantly affected during an interaction time \cite{Eijnden}. Moreover, in the absence of a physical model connecting a mode $m$ to a mode $m'$, we can simply assume that correlation only exists when $m=m'$. This simplification is likely incorrect for ULF waves within plumes, but offers nonetheless a closure to the above equation and a lower bound to the diffusion equation, i.e., accounting for non-zero correlations between $m$ and $m'$ terms will enhance diffusion. 

We proceed to compute the correlator by first taking the product of $c_m$ and $c_m^*$:
\begin{eqnarray}\label{product_c_m}
    c_m(t')c_m^*(t)&=&\left(\sum_q \frac{\delta E_{\varphi, q}}{2\pi}(\delta_{m,q}+2\varepsilon_{m,q}\psi_{q-m})\right)\left(\sum_p \frac{\delta E_{\varphi, p}^*}{2\pi}(\delta_{m,p}+2\varepsilon_{m,p}\psi_{p-m})\right) \nonumber \\
    &=&\sum_q \sum_p \frac{\delta E_{\varphi, q}\delta E_{\varphi, p}^*}{4\pi^2} (\delta_{m,q}+2\varepsilon_{m,q}\psi_{q-m})(\delta_{m,p}+2\varepsilon_{m,p}\psi_{p-m})
\end{eqnarray}
and then take the drift and time average to find: 
\begin{eqnarray}\label{product_average}
    \langle c_m (t')c_m^*(t)\rangle_\varphi&=&\sum_q \sum_p \bigg{\langle}\frac{\delta E_{\varphi, q}\delta E_{\varphi, p}^*}{4\pi^2}\bigg{\rangle}_\varphi (\delta_{m,q}+2\varepsilon_{m,q}\psi_{q-m})(\delta_{m,p}+2\varepsilon_{m,p}\psi_{p-m}) \nonumber \\ 
    &=&\delta(t'-t)\sum_q \sum_p \bigg{\langle}\frac{\delta E_{\varphi, q}^2}{4\pi^2}\bigg{\rangle}_\varphi \delta_{q,p}(\delta_{m,q}+2\varepsilon_{m,q}\psi_{q-m})(\delta_{m,p}+2\varepsilon_{m,p}\psi_{p-m}) \nonumber  \\
    &=&\delta(t'-t)\sum_q\bigg{\langle}\frac{\delta E_{\varphi, q}^2}{4\pi^2}\bigg{\rangle}_\varphi(\delta_{m,q}+2\varepsilon_{m,q}\psi_{q-m})(\delta_{m,q}+2\varepsilon_{m,q}\psi_{q-m}) \nonumber \\    
    &=&\delta(t'-t)\sum_{q}\bigg{\langle}\frac{\delta E_{\varphi, q}^2}{4\pi^2}\bigg{\rangle}_\varphi(\delta_{m,q}+4\varepsilon_{m,q}\psi_{q-m}^2)
\end{eqnarray}
where in the last line we took advantage of the following identities: 
\begin{itemize}
    \item $\sum_q \delta_{m,q}^2=1$ if $m=q$ and zero otherwise.
    \item $\sum_q \delta_{m,q} \varepsilon_{m,q}=0$ for all combinations of $m$ and $q$.
    \item $\sum_q \varepsilon_{m,q}^2=\sum_q \varepsilon_{m,q}=1$ if $m\neq q$ and zero otherwise. 
\end{itemize}
Thus, inserting the correlator $\big{\langle} c_{m}(t')c_{m}^*(t) \big{\rangle}_\varphi$ in the above Equation we find the quasi-linear radial diffusion equation for MLT localised waves: 
%%%%%%%%%%%%%%%%%%%%%%%%%%%%%%%%
\begin{eqnarray}
   \frac{\partial g_0}{\partial t}&=& \frac{1}{B_E^2}\frac{r^2}{R_E^2}\frac{\partial}{\partial r} \bigg{[} \frac{r^2}{R_E^2} \sum_m \sum_{m'}\bigg{\langle}\frac{\delta E_{\varphi, m'}^2}{4\pi^2}\bigg{\rangle}_\varphi(\delta_{m,m'}+4\varepsilon_{m,m'}\psi_{m'-m}^2)\frac{\partial g_0}{\partial r} \bigg{]} \nonumber \\
    &=& L^2 \frac{\partial}{\partial L}\bigg{(}\frac{D_{LL}}{L^2}\frac{\partial g_0}{\partial L}\bigg{)} 
\end{eqnarray}
%%%%%%%%%%%%%%%%%%%%%%%%%%%%%%%%
with the radial diffusion coefficient given by: 
%%%%%%%%%%%%%%%%%%%%%%%%%%%%%%%%
\begin{eqnarray}\label{DLL_final}
    D_{LL}&=& \sum_m \sum_{m'}  \frac{L^4}{B_E^2R_E^2} \bigg{\langle}\frac{\delta E_{\varphi, m'}^2}{4\pi^2}\bigg{\rangle}_\varphi (\delta_{m,m'}+4\varepsilon_{m,m'})\psi_{m'-m}^2 \nonumber \\ 
    &=& \frac{L^4}{B_E^2R_E^2}\sum_m \bigg{\langle}\frac{\delta E_{\varphi, m}^2}{4\pi^2}\bigg{\rangle}_\varphi +  \frac{L^4}{B_E^2R_E^2}\sum_{m'}\sum_{m} \bigg{\langle}\frac{\delta E_{\varphi, m'}^2}{4\pi^2}\bigg{\rangle}_\varphi 4\varepsilon_{m'm}\psi^2_{m-m'}\nonumber \\
    &=& \frac{L^4}{B_E^2R_E^2}\sum_m \bigg{\langle}\frac{\delta E_{\varphi, m}^2}{4\pi^2}\bigg{\rangle}_\varphi +  \frac{L^4}{B_E^2R_E^2}\sum_{m'}\bigg{\langle}\frac{\delta E_{\varphi, m'}^2}{4\pi^2}\bigg{\rangle}_\varphi\sum_{m}  4\varepsilon_{m'm}\psi^2_{m-m'}\nonumber \\
     &=&\frac{L^4}{B_E^2R_E^2}\sum_m \bigg{\langle}\frac{\delta E_{\varphi, m}^2}{4\pi^2}\bigg{\rangle}_\varphi  \left(1+4\sum_{q} \varepsilon_{m+q,m}\psi_{q}^2 \right)
\end{eqnarray}
%%%%%%%%%%%%%%%%%%%%%%%%%%%%%%%%
where we reversed the sums for $m$ and $m'$ between the second and third line, inverted the indices $m\leftrightarrow m'$ between the third and fourth line, wrote the sum in terms of $q=m'-m$ and took advantage of the identity $\psi_{q}=\psi_{-q}$. The quasi-linear radial diffusion in Equation \ref{DLL_final} for MLT localised ULF waves is almost identical to the radial diffusion coefficient for the special case of ULF fields uniformly distributed along MLT. The difference is the positive definite factor of 
\begin{equation}
    1+4\sum_{q} \varepsilon_{m+q,m}\psi_{q}^2(\kappa),
\end{equation} 
which is a function of the parameter $\kappa$ and originates from the von Mises distribution. If we set $\kappa=0$, we recover the radial diffusion for ULF waves uniformly distributed in magnetic local time: $D_{LL}=\frac{L^4}{B_E^2R_E^2}\sum_m \bigg{\langle}\frac{\delta E_{\varphi, m}^2}{4\pi^2}\bigg{\rangle}_\varphi$ since $\psi_q(\kappa=0) \neq 0$ only for $q=0$ but $\varepsilon_{m+q, q}=0$ if $q\neq 0$. But if $\kappa \neq 0$, and as the ULF waves become more localised, the additional coefficient increases the value of radial diffusion monotonically. 

 \begin{figure}
  %\centering
  % \includegraphics[width=0.383\textwidth]{Fig1a.eps}
  \includegraphics[width=0.5\textwidth]{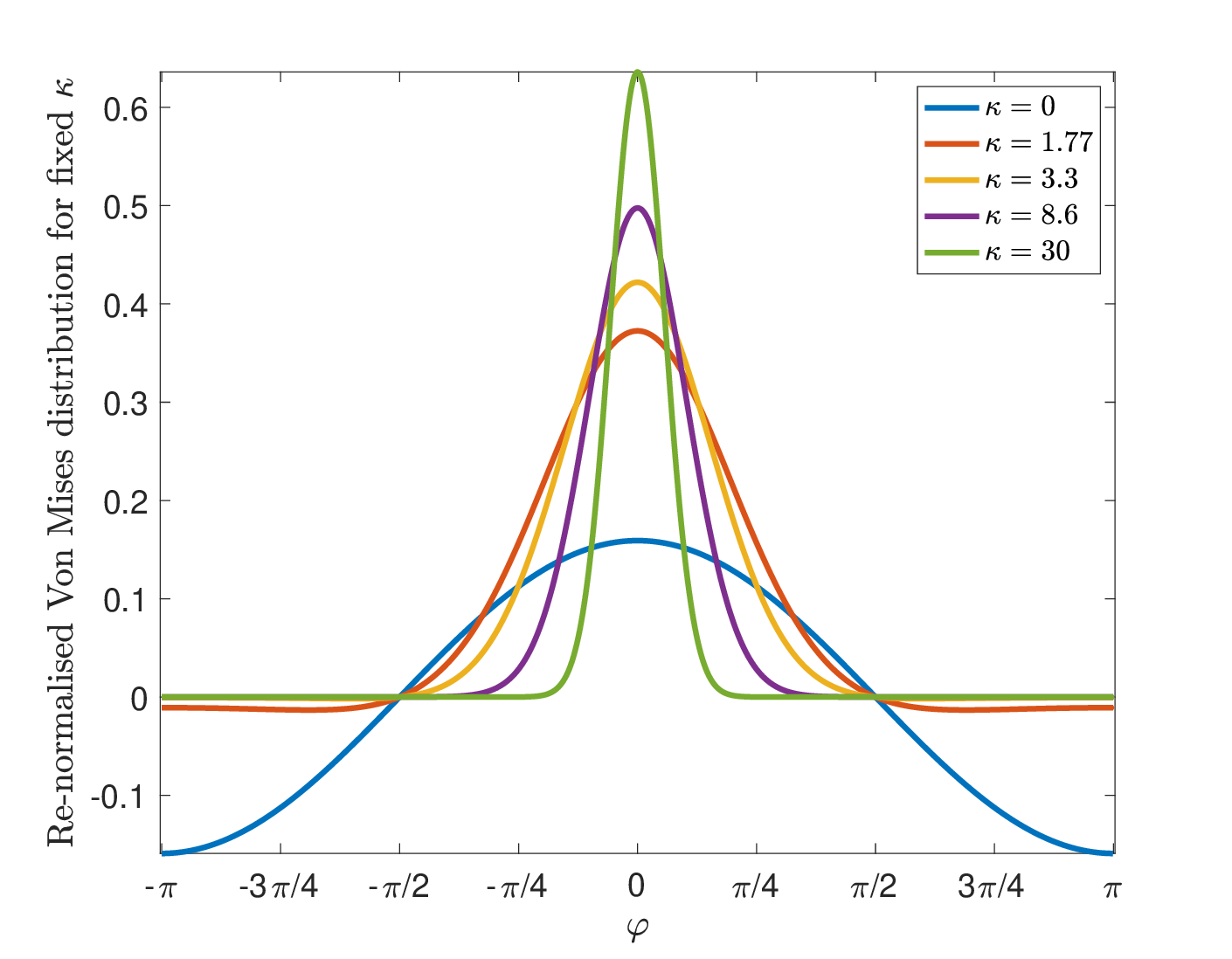}
    \includegraphics[width=0.5\textwidth]{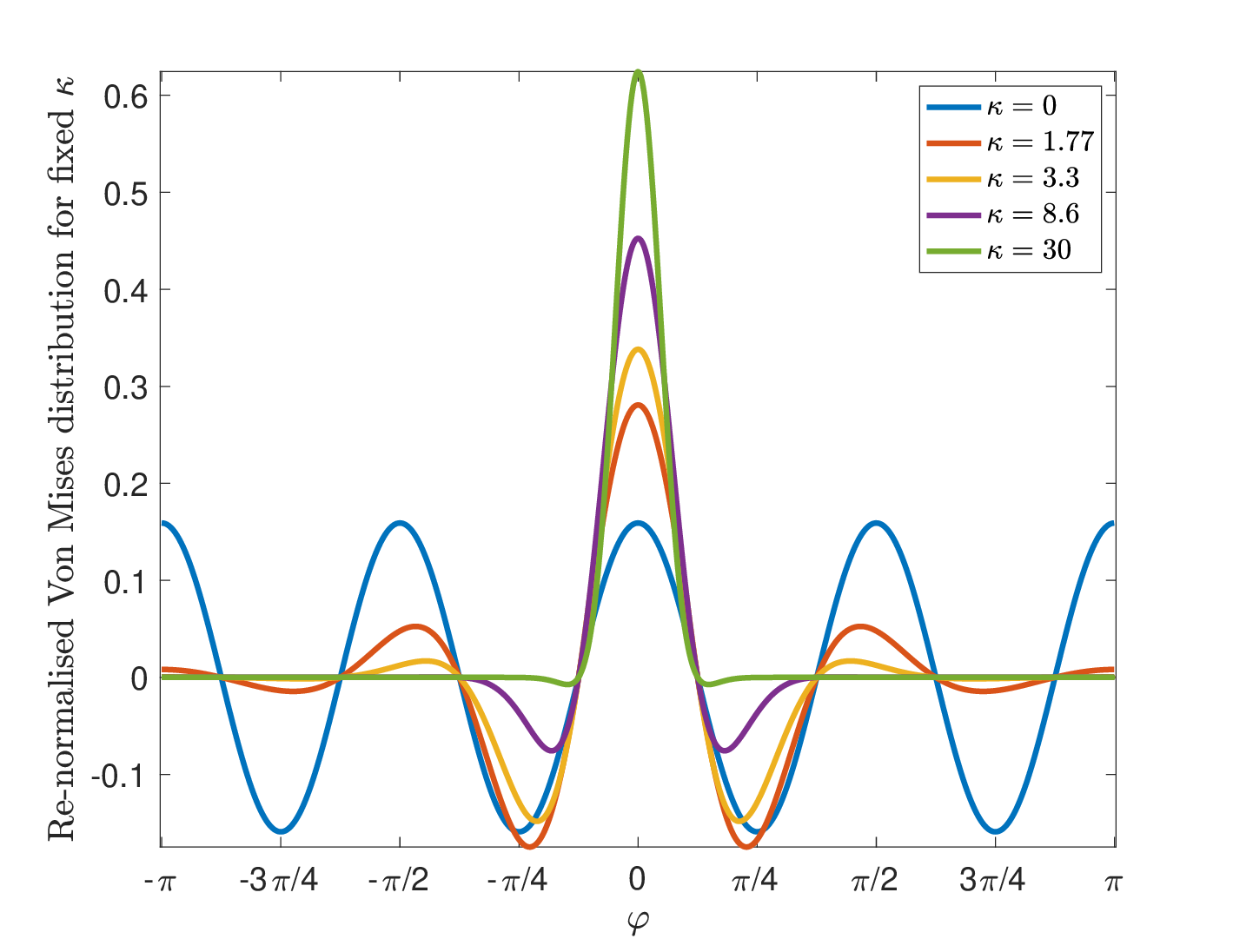}
 \caption{MLT dependence of ULF waves with $m=1$ (left panel) and $m=4$ (right panel) with a von Mises envelopes $f(\varphi; \kappa)$, $\kappa$ values of $[0, 1.77, 3.3, 8.6, 30]$, and amplitude rescaling by a fixed root mean square fluctuation $f(\varphi; \kappa)\longrightarrow \frac{f(\varphi;\kappa)}{\sqrt{1+2\sum_{q, q\neq m}\psi_{q-m}}}$ given by Equation (\ref{renormalisation}). The selected $\kappa$ values correspond to cases where the envelope confines ULF waves to approximately 100, 75, 50, 30 and 15 percent of the drift orbit.}
  \label{Fig_3}
 \end{figure}

\begin{figure}
\centering
 \includegraphics[width=0.85\textwidth]{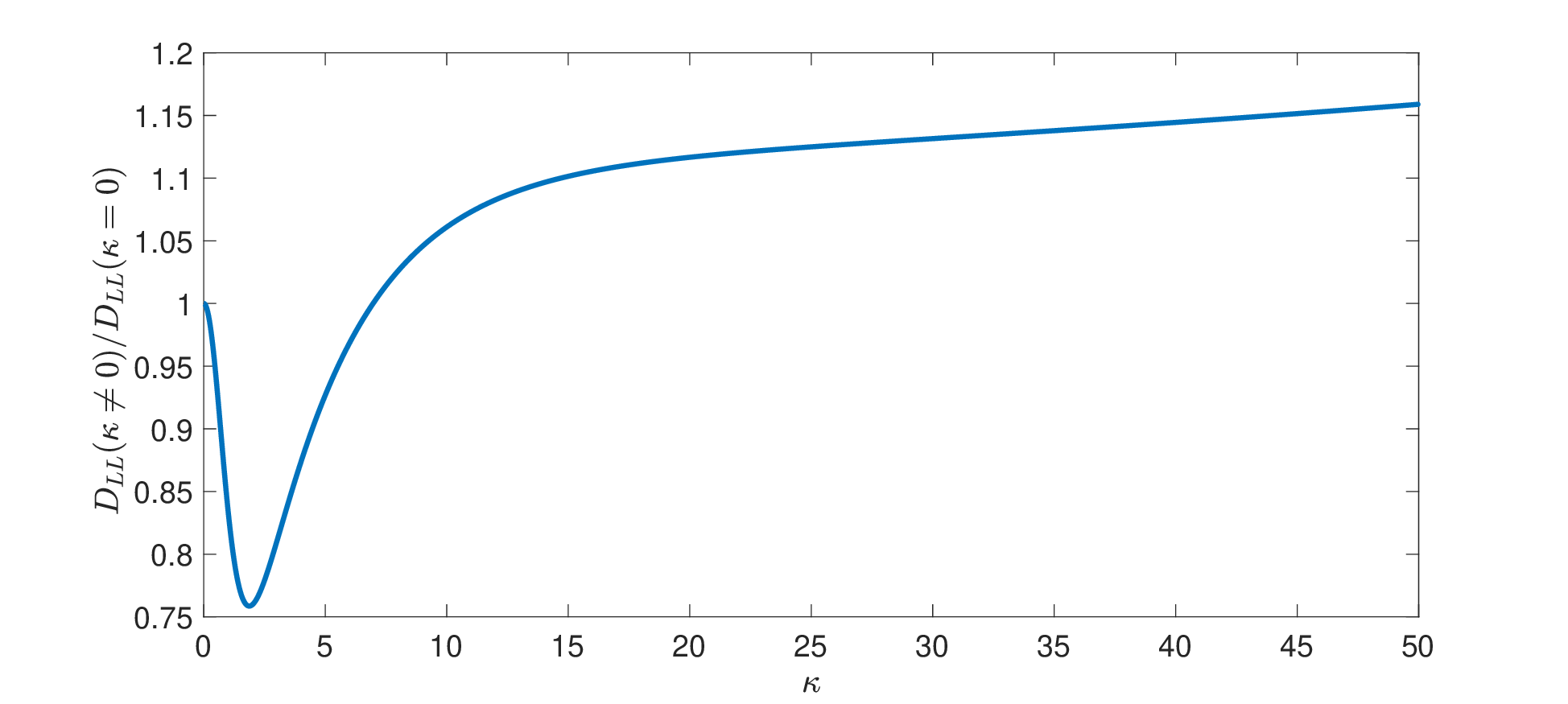}
  \includegraphics[width=0.85\textwidth]{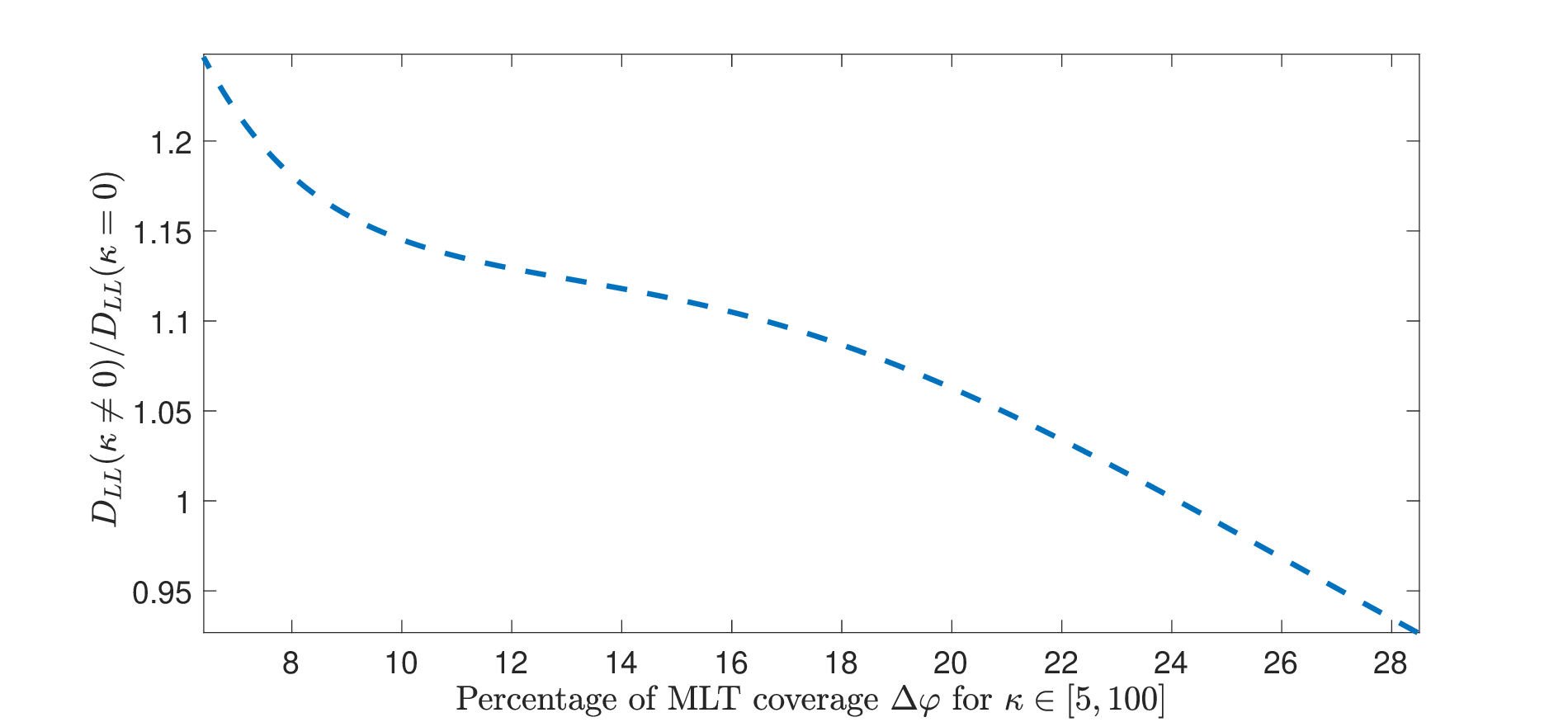}
 \caption{Top panel (full line) shows the relative change of the radial diffusion coefficient $D_{LL}$ as a function of the $\kappa$ parameter. The parameter $\kappa$ determines the spread of ULF MLT range. The higher the value of $\kappa$ the more localised are the ULF waves. The diffusion coefficient is re-scaled by Equation (\ref{renormalisation}) to ensure that a particle encounters the same root-mean-square fluctuation during a drift orbit.  The relative increase in $D_{LL}$ is zero when $\kappa=0$, and grows by about 15\% when $\kappa>30$. The bottom panel (dashed line) shows the relative change of the radial diffusion coefficient $D_{LL}$ as a function of the percentage of MLT coverage for $\kappa=5$ (30\%) and $\kappa=100$ (6\%).}
 \label{Fig_4}
 \end{figure}

%Even though the von Mises distribution function is normalised the maximum wave amplitude grows by a factor of 10 between $\kappa=0$ and $\kappa=30$, as shown in Figure (\ref{Fig_2}). 
Even though the von Mises distribution function is normalised, the maximum wave amplitude grows by a factor of 10 between $\kappa=0$ and $\kappa=30$. We therefore need to rescale the wave amplitude in a manner that is not affected by the $\kappa$ parameter. In order to do so, we compute the root mean square fluctuation $\sqrt{\langle\delta E_\varphi^2\rangle}$, where the brackets denote the drift average, and add the additional requirement that a particle samples the same root mean square fluctuation on a drift orbit. The computation of ${\langle\delta E_\varphi^2\rangle}$ is presented in the \ref{Appendix_Fourier}. We find that for a single mode $m$, the root mean square fluctuation $ \langle \delta E_{m,\varphi}^2\rangle$ needs to be rescaled according to: 
%%%%%%%
\begin{equation}\label{rescale_1}
    \frac{\langle \widetilde{{\delta E}}_{\varphi, m}^2\rangle}{4\pi^2} = \frac{\langle \delta E_{\varphi}^2\rangle}{1+2\sum_{p, p\neq m}\psi_{p-m}}.
\end{equation}
%%%%%%%
The impact of this rescaling of the ULF wave is shown in Figure \ref{Fig_3} for $m=1$ and $m=4$ modes. The blue lines in both panels show the ULF fluctuations that are homogeneous in MLT. The radial diffusion coefficient re-scaled to account for fixed root mean square fluctuations along a drift orbit is given by: 
%%%%%%%%%%%%%%%%%%%%%%%%%%%%%%%%
\begin{eqnarray}\label{DLL_final_final}
    D_{LL}(\kappa\neq0) =\frac{L^4}{B_E^2R_E^2} \big{\langle}{\delta E_{\varphi}^2}\big{\rangle}_\varphi \sum_m \left(\frac{1+4\sum_{q} \varepsilon_{m+q,m}\psi_{q}^2}{{1+2\sum_{q}\varepsilon_{m+q,m}\psi_q}}\right)
\end{eqnarray}
%%%%%%%%%%%%%%%%%%%%%%%%%%%%%%%%
whereas the diffusion coefficient for MLT homogeneous fluctuations is given by 
%%%%%%%%%%%%%%%%%%%%%%%%%%%%%%%%
\begin{eqnarray}\label{DLL_final_final}
    D_{LL}(\kappa=0) =\frac{L^4}{B_E^2R_E^2} \big{\langle}{\delta E_{\varphi}^2}\big{\rangle}_\varphi.
\end{eqnarray}
%%%%%%%%%%%%%%%%%%%%%%%%%%%%%%%%
\noindent In Figure \ref{Fig_4} we show the ratio of the radial diffusion coefficient when $\kappa\neq 0$ to the one when $\kappa=0$ for fixed root mean square fluctuation $ \langle \delta E_{\varphi}^2\rangle$ and for a fixed $m$: 
%%%%%%%
\begin{equation}
    \frac{D_{LL}(\kappa\neq0)}{D_{LL}(\kappa=0)}=\frac{1+4\sum_{q} \varepsilon_{q+m,m}\psi_q^2(\kappa)}{1+2\sum_{q}\varepsilon_{q+m,m}\psi_q(\kappa)}.
\end{equation}
%%%%%%%
The relative increase in $D_{LL}$ due to wave power in the azimuthal wave number $m$ is given by $4\sum_{q} \varepsilon_{m+q, m}\psi_q^2$ and is equal to zero when $\kappa=0$, yet increases by about 16\% even when the drift orbit encounters ULF waves for less than 20\% of its orbit, or equivalently for $\kappa>25$. This result may seem surprising, as it suggests that even when ULF waves are encountered over a small fraction of the drift orbit, they can still induce radial diffusion comparable to cases where the ULF wave is sampled across all magnetic local times. In the next section we provide mathematical and physical explanation to this result. 

\subsection{Explanation for enhanced diffusion despite narrowly localised waves.} \label{explanation}
Our analysis indicates that while quasi-linear radial diffusion does not retain information about the magnetic local time, the radial diffusion due to narrowly localised ULF waves is more efficient. This non-trivial result requires a physical explanation. The answer can be found by solving the linear perturbed part of the distribution function $g_{m}$ for a single wave mode of frequency $\omega_{m}$ and growth or damping rate $\gamma_{m}$: 
%%%%%%%%%%%%%%%%%%%%%%%%%%%%%%%%
\begin{equation}
\label{Eq:single_wave_mode_1}
    \delta E_{\varphi} =  c_{m}(t) e^{im\varphi}
\end{equation}
%%%%%%%%%%%%%%%%%%%%%%%%%%%%%%%%
with 
%%%%%%%%%%%%%%%%%%%%%%%%%%%%%%%%
\begin{equation}
\label{Eq:single_wave_mode_2}
c_m=\sum_{m'} \frac{\widetilde{{\delta E}}_{\varphi, m'}}{2 \pi} \left(\delta_{m, m'}+2\varepsilon_{m,m'}\psi_{m'-m}\right) e^{(i \omega_{m'}+\gamma_{m'})t},
\end{equation}
%%%%%%%%%%%%%%%%%%%%%%%%%%%%%%%%
We also require that $\gamma_{m'}\ll \omega_{m'}$, such that a drifting particle transits the wave before it damps or grows to large-amplitudes where nonlinear effects would become significant. Inserting Equation (\ref{Eq:single_wave_mode_1}) and (\ref{Eq:single_wave_mode_2}) into Equation (\ref{linear_solution}) gives: 
%%%%%%%%%%%%%%%%%%%%%%%%%%%%%%%%
\begin{eqnarray}
\label{EQ:multiple_resonances}
   \delta g_{m}(r,t)&=&\sum_{m'} \frac{\widetilde{{\delta E}}_{\varphi, m'}}{2 \pi} \left(\delta_{m, m'}+2\varepsilon_{m,m'}\psi_{m'-m}\right)e^{i m\Omega_d t}\int_0^t  \mathrm{d}t'\   e^{i(\omega_{m'}-m\Omega_d-i\gamma_{m'})t'}\frac{\partial g_0}{\partial r}   \nonumber \nonumber \\ 
   &=&-i\sum_{m'} \frac{\widetilde{{\delta E}}_{\varphi, m'}}{2 \pi} \left(\delta_{m, m'}+2\varepsilon_{m,m'}\psi_{m'-m}\right)e^{i m\Omega_d t}\left(\frac{e^{i(\omega_{m'}-m\Omega_d-i\gamma_{m'})t}-1}{\omega_{m'}-m\Omega_d-i\gamma_{m'}}\right)\frac{\partial g_0}{\partial r} \nonumber  \\
    &=&-i\underbrace{\frac{\widetilde{{\delta E}}_{\varphi, m}}{2 \pi}\left(\frac{e^{i(\omega_{m}-m\Omega_d-i\gamma_{m})t}-1}{\omega_{m}-m\Omega_d-i\gamma_{m}}\right)}_{\textrm{\small{Drift Resonance for $\omega_m\simeq m\Omega_d$}}}\frac{\partial g_0}{\partial r}e^{i m\Omega_d t} \nonumber \\
    &-&i\underbrace{\sum_{m'\neq m}\psi_{m'-m}\frac{\widetilde{{\delta E}}_{\varphi, m'}}{ \pi}\left(\frac{e^{i(\omega_{m'}-m\Omega_d-i\gamma_{m'})t}-1}{\omega_{m'}-m\Omega_d-i\gamma_{m'}}\right)}_{\textrm{\small{Additional drift resonances for  $\omega_{m'}\simeq m\Omega_d$}}}\frac{\partial g_0}{\partial r}e^{i m\Omega_d t} 
\end{eqnarray}
%%%%%%%%%%%%%%%%%%%%%%%%%%%%%%%%
The mathematical answer to the enhanced radial transport due to narrowly localised MLT waves is in Equation (\ref{EQ:multiple_resonances}). We recover the drift resonance $\omega_m\simeq m\Omega_d$ one would find for MLT homogeneous waves \cite{osmane2023radial}. But if the waves are MLT localised, we see the appearance of additional resonant terms when $\omega_{m'} \simeq m\Omega_d$ for $m'\neq m$. If $m=m'+n$, the additional resonance can be written as $\omega_{m'}\simeq (m'+n) \Omega_d$ for $n\neq0$. These new drift resonance terms are weighted by the Bessel function ratio $\psi_{m'-m}$, which only become significant when $\kappa \neq 0$. Thus, when the drift resonant perturbations are inserted in the quasi-linear diffusion equation, terms with $n\neq 0$ result in additional radial diffusion terms. While these terms also have to be re-scaled according to Equation (\ref{rescale_1}) in order to make sure that the particles sample the same root-mean-square fluctuations along a drift orbit, they are positive definite additional sources for radial diffusion. 

Mathematically, the above description should be satisfying to the reader since it shows that additional drift resonances are now present when ULF waves are localised in MLT. But physically why is it that confining ULF waves to a narrow range of azimuthal angles results in increased resonant interactions? From the particle's perspective, the sampled wave field fluctuates in time but remains localised in the azimuthal angle $\varphi$. The more confined the wave is in the azimuthal direction, the larger the corresponding range of $m$ values that the particle samples. As a result, the particle does not encounter a wave with distinct peaks and troughs, but rather a localized structure confined to part of its drift orbit (as illustrated in Figure \ref{Fig_3}).
Drift resonance can still occur with various harmonics $m'\neq m$ if a collection of particles with drift period $\Omega_d$ experiences a localised but net electric field. Furthermore, since the MLT regions without ULF do not contribute or affect transport radial transport is not precluded even if ULF waves are encountered only for a small portion of the drift trajectory. And thus, even when ULF waves are confined to a narrow range of MLT, radial transport can occur as long as repeated encounters with the ULF region lead to the sampling of a net electric field \textit{on average}, that is for a collection of particles. We therefore find that radial diffusion remains comparably efficient when particles sample the same root-mean-square ULF fluctuations over a small fraction of the drift orbit, as it does over the entire drift orbit, due to the presence of additional drift resonances.

\section{Conclusions}\label{conclusion}
For decades, it has been well established that ULF waves in the Earth’s magnetosphere exhibit significant localization in magnetic local time (MLT), with clear asymmetries observed across different sectors of the magnetosphere \cite{Gupta75,Anderson90,Vennestrom99,Liu09,Pahud09,Rae12,Kokubun13,Takahashi15,Takahashi16,Murphy20,Sarris22,Yan23}. Despite this, quasi-linear radial diffusion coefficients, derived analytically \cite{Falthammar65, Elkington03, Fei06, Lejosne19, osmane2023radial}, typically assume that ULF waves are uniformly encountered across all MLT sectors by magnetically trapped particles. This assumption overlooks the well-documented MLT-dependent variations in ULF wave power and structure. 

In this communication, we present, for the first time, a quasi-linear radial diffusion coefficient for ULF wave modes localized in MLT. Our results demonstrate that while the quasi-linear distribution function is averaged over MLT and drift orbits, narrow MLT localization of ULF waves can significantly impact particle transport. We find that when ULF waves cover more than 30\% of the MLT, the efficiency of radial diffusion is comparable to scenarios where ULF waves are uniformly distributed across all MLT sectors. However, ULF waves encountered on less than 10\% of the drift orbit lead to an enhanced radial diffusion, increasing by 10\% to 25\%. While these enhancements may seem modest when accounting for other sources of variability in empirically derived radial diffusion coefficients \cite{Thompson_variability_DLL, Sandhu21}, they indicate that even narrowly localized ULF waves can be an efficient driver of radial transport in planetary magnetospheres.

While our derivation is informative, and indicative of additional drift resonance when ULF waves are MLT localized, it has several limitations that must be addressed to fully account for the role of ULF waves in radial diffusion. Perhaps the most significant limitation is that the ULF field used in our derivation is an electrostatic poloidal field superposed on a dipolar field. In solar wind-driven magnetospheres, a more realistic representation of ULF waves should include both toroidal and poloidal electromagnetic components (see, e.g., \citeA{Murphy20} and references therein) and static non-dipolar contributions \cite{Fei06, Cunningham16}. Recent numerical modelling works have shown that the 3D inhomogeneous nature of the magnetosphere has a crucial impact on the resulting ULF waves \cite{Degeling2018, WrightElsden2020}. These works show that azimuthal variations in the plasma mass density, for example from a plasmaspheric plume, fundamentally alter the polarisation of ULF waves, as well as their spatial extents. The classical 1D theory of ULF wave coupling describing field line resonance (FLR) \cite{Southwood74, ChenHasegawa74} assuming azimuthal symmetry is no longer suitable in such a 3D varying plasma \cite{WrightElsden2016}. This study represents the first attempt to bring the radial diffusion theory in line with some of the inhomogeneous aspects of ULF waves in observations and simulations. 

A second set of limitations, which also broadly applies to other quasi-linear derivations, is that we focused exclusively on particles with 90-degree pitch angles and assumed that the various azimuthal modes $m$ and $m'$ (with $m' \neq m$) are short, uncorrelated impulses. While the former limitation can be easily addressed for arbitrary pitch angles \cite{Schulz74}, the latter assumption requires a statistical model of ULF waves to account for the correlations between different azimuthal wave modes. A third limitation is that plumes and regions with radially localized ULF waves are time-dependent, and this variability can occur on timescales comparable to drift orbits. For example, the plume structure can change significantly over a few hours \cite{Goldstein14}, and the radial structure of ULF waves is highly variable in time due to solar wind dynamic pressure variations \cite{Claudepierre10} and Alfv\'{e}n wave phase-mixing \cite{MannWright95}. We do not address this time dependence in the current communication, but a first step in generalizing the model would be to quantify the impact of a time-varying $\kappa$ parameter in the ULF wave field. Addressing these limitations within quasi-linear diffusive and non-diffusive frameworks in future work will provide a more comprehensive understanding of the complex interactions between ULF waves and energetic particles, ultimately leading to more realistic models of radial diffusion in planetary magnetospheres.

%\newpage
\appendix
\section{Fourier Representation and Spectra}\label{Appendix_Fourier}
In this section we seek a Fourier representation to the MLT localised poloidal electric field in Equation (\ref{electric_field_eq2})
%%%%%%%
\begin{eqnarray}
{\delta E_\varphi}&=& f(\varphi; \kappa) \sum_m \widetilde{{\delta E}}_{\varphi, m}e^{im\varphi}, 
\end{eqnarray}
%%%%%%%
with the envelope given by Equation (\ref{Eq:vonMises}) and the variable $\widetilde{{\delta E}}_{\varphi, m}$ containing the radial dependence. 
We want instead to have an equation for the electric of the form: 
%%%%%%%
\begin{equation}
    \delta E_\varphi(r, \varphi, t)=\sum_k c_k e^{i k\varphi},
\end{equation}
%%%%%%%
with the Fourier coefficients $c_k=c_k(r,t)$ given by
%%%%%%%
\begin{eqnarray}
  c_k&=& \frac{1}{2\pi} \int_{-\pi}^\pi \mathrm{d}\varphi \ \delta E_\varphi e^{-ik\varphi}  \nonumber \\ 
  &=& \frac{1}{2\pi} \sum_m \widetilde{{\delta E}}_{\varphi, m}  \int_{-\pi}^\pi \mathrm{d}\varphi \ f(\varphi; \kappa) e^{i (m-k)\varphi} 
\end{eqnarray}
%%%%%%%
The above integral is easily solved with the Fourier convolution theorem and gives the following result for the Fourier coefficients: 
%%%%%%%
\begin{eqnarray}
    c_m=\sum_{m'} \frac{\widetilde{{\delta E}}_{\varphi, m'}}{2 \pi} \left(\delta_{m, m'}+2\varepsilon_{m,m'}\psi_{m'-m}\right),
\end{eqnarray}
%%%%%%%
where we used the Kronecker delta  $\delta_{m, m'}$ for indices $m$ and $m'$, it's complement $\varepsilon_{m,m'}=1-\delta_{m,m'}$ and the function $\psi_{m'-m}$ as the ratio of the modified Bessel functions $I_{n}$ of order $m'-m$ and zero. The Fourier representation for the poloidal electric field can therefore be written as: 
%%%%%%%
\begin{eqnarray}
    \delta E_\varphi= \sum_m\sum_{m'} \frac{\widetilde{{\delta E}}_{\varphi, m'}}{2 \pi} \left(\delta_{m, m'}+2\varepsilon_{m,m'}\psi_{m'-m}\right) e^{im\varphi}
\end{eqnarray}
%%%%%%%
 When $\kappa=0$, $\varepsilon_{m,m'} \psi_{m'-m}=0$ for any combinations of $m$ and $m'$ and the Fourier coefficients reduce to $c_m=\frac{\widetilde{{\delta E}}_{\varphi, m}}{2 \pi}$, as it should. 

Let's now compute the Fourier spectra $\langle \delta E_{\varphi}\delta E_{\varphi}^*\rangle_\varphi$ in terms of the time and space average given by Equation (\ref{averaging_procedure}). The first step is to write the product of $ \delta E_\varphi$ to itself:
%%%%%%%
\begin{eqnarray}\label{Eq:product_spectra}
    \delta E_{\varphi} \delta E_{\varphi}^*&=&\left(\sum_m\sum_{m'} \frac{\widetilde{{\delta E}}_{\varphi, m'}}{2 \pi} \left(\delta_{m, m'}+2\varepsilon_{m,m'}\psi_{m'-m}\right) e^{im\varphi}\right)\left(\sum_p\sum_{q} \frac{\widetilde{{\delta E}}_{\varphi, q}^*}{2 \pi} \left(\delta_{p, q}+2\varepsilon_{p,q}\psi_{q-p}\right) e^{-iq\varphi}\right) \nonumber \\
&=&\sum_m\sum_{m'}\sum_p\sum_{q}\frac{\widetilde{{\delta E}}_{\varphi, m'}}{2 \pi} \frac{\widetilde{{\delta E}}_{\varphi, q}^*}{2 \pi}e^{i(m-q)\varphi}\left(\delta_{m, m'}+2\varepsilon_{m,m'}\psi_{m'-m}\right)\left(\delta_{p, q}+2\varepsilon_{p,q}\psi_{q-p}\right)
\end{eqnarray}
%%%%%%%
After averaging in phase and time, and assuming the short time correlation assumption, i.e., $\langle \widetilde{{\delta E}}_{\varphi, m'}\widetilde{{\delta E}}_{\varphi, q}^*\rangle = \langle \widetilde{{\delta E}}_{\varphi, m'}^2\rangle \delta_{m', q}$
Equation (\ref{Eq:product_spectra}) yields: 
%%%%%%%
\begin{eqnarray}\label{Eq:product_spectra2}
    \langle \delta E_{\varphi}^2\rangle &=&\frac{1}{4\pi^2}\sum_m\sum_{m'}\sum_p\sum_{q} \langle \widetilde{{\delta E}}_{\varphi, m'}^2\rangle \delta_{q, m'} \delta_{m, q} \left(\delta_{m, m'}+2\varepsilon_{m,m'}\psi_{m'-m}\right)\left(\delta_{p, q}+2\varepsilon_{p,q}\psi_{q-p}\right)\nonumber \\ 
    &=&\frac{1}{4\pi^2}\sum_m\sum_{q}\sum_p \langle \widetilde{{\delta E}}_{\varphi, q}^2\rangle  \delta_{m, q}\left(\delta_{m, q}+2\varepsilon_{m,q}\psi_{q-m}\right)\left(\delta_{p, q}+2\varepsilon_{p,q}\psi_{q-p}\right)\nonumber \\
%    &=&\frac{1}{4\pi^2}\sum_m\sum_p \langle \widetilde{{\delta E}}_{\varphi, m}^2\rangle \left(\delta_{m, m}+2\psi_{m-m}\right)\left(\delta_{p, m}+2\psi_{m-p}\right) \nonumber \\ 
    &=&\frac{1}{4\pi^2}\sum_m\sum_p \langle \widetilde{{\delta E}}_{\varphi, m}^2\rangle \left(\delta_{p, m}+2\varepsilon_{m,p}\psi_{m-p}\right) \nonumber\\
    &=&\frac{1}{4\pi^2}\sum_m \langle \widetilde{{\delta E}}_{\varphi, m}^2\rangle\left(1+2\sum_{p, p\neq m} \psi_{p-m}\right)   
    \end{eqnarray}

\begin{figure}
\centering
\includegraphics[width=.66\textwidth]{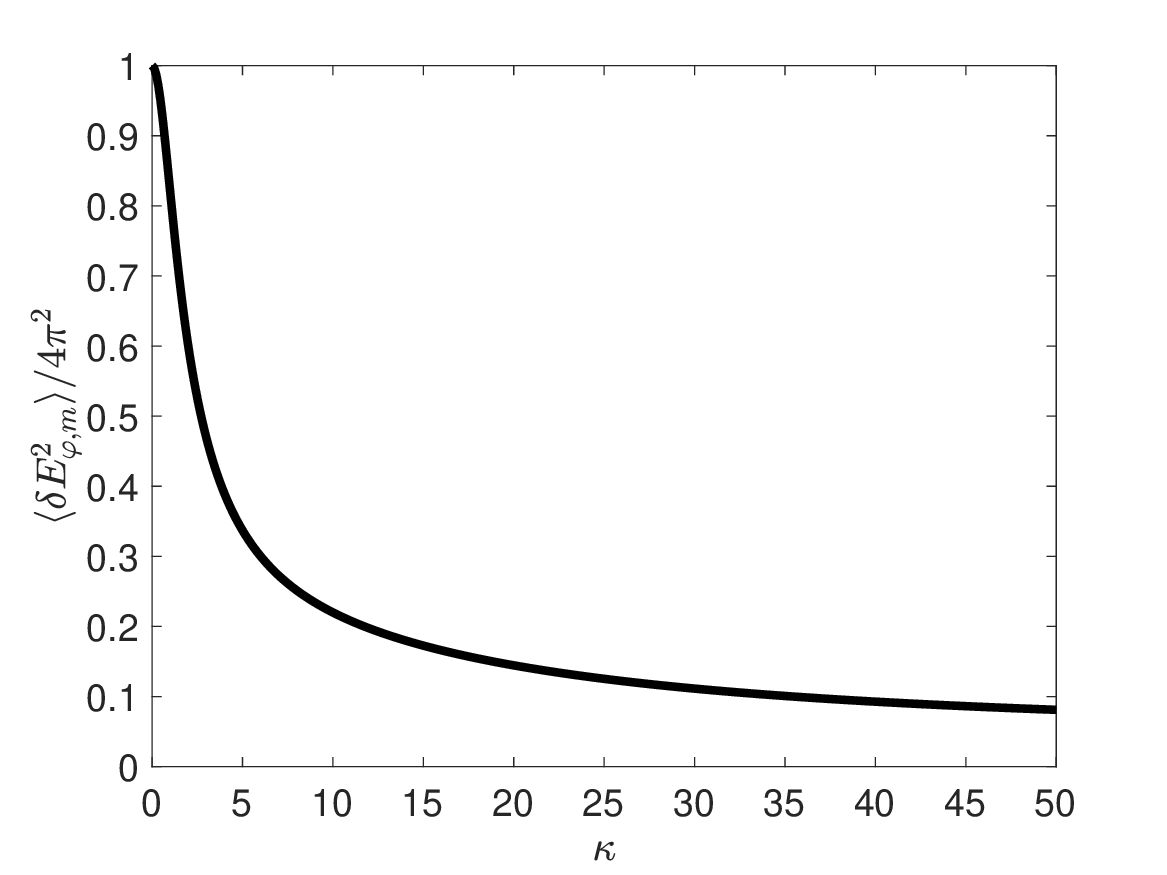}
        \caption{Dependence of $m=1$ ULF wave power for a given value of $\kappa$ when we require that the particle encounters a fixed root mean square fluctuation on the drift orbit. See Equation (\ref{renormalisation}) and associated text.}
        \label{fig:4}
\end{figure}

In Figure (\ref{fig:4}) we plot $\frac{1}{4\pi^2} \langle \widetilde{{\delta E}}_{\varphi, m}^2\rangle$ for a single mode $m$ as a function $\kappa$ for a fixed root mean square fluctuation $ \langle \delta E_{\varphi}^2\rangle=1$: 
%%%%%%%
\begin{equation}\label{renormalisation}
    \frac{\langle \widetilde{{\delta E}}_{\varphi, m}^2\rangle}{4\pi^2} = \frac{\langle \delta E_{\varphi}^2\rangle}{1+2\sum_{p, p\neq m}\psi_{p-m}}.
\end{equation}
%%%%%%%
Figure  (\ref{fig:4}) shows that the wave power for a given azimuthal wave number $m$ must decrease by a factor of 10 for large values of $\kappa\geq 25$ in order to account for fixed root mean square fluctuations sampled on a drift trajectory.

\acknowledgments
Support for AO was provided by the University of Helsinki and the Research Council of Finland profiling action Matter and Materials (grant \# 318913). OA would like to acknowledge financial support from the University of Birmingham, the University of Exeter, and also from the United Kingdom Research and Innovation (UKRI) Natural Environment Research Council (NERC) Independent Research Fellowship NE/V013963/1 and NE/V013963/2. The work of LT is funded by the European Union (ERC grant WAVESTORMS - 101124500). Views and opinions expressed are however those of the author(s) only and do not necessarily reflect those of the European Union or the European Research
Council Executive Agency. Neither the European Union nor the granting authority can be held responsible for them. LT also acknowledges support from the Research Council of Finland (grant number 322544).

\bibliography{agusample}

\end{document}